\documentclass[a4paper]{jpconf}
\usepackage{graphicx}
\usepackage{amssymb}

\def\al{\alpha} \def\be{\beta}  \def\de{\delta}
   
   \def\ka{\kappa}
\def\la{\lambda}   
\def\si{\sigma}

\def\La{\Lambda}   
  \def\mn{{\mu\nu}}

 \def\frac#1#2{{\textstyle{{#1}\over
{#2}}}} 
\def\lsim{\mathrel{\rlap{\lower4pt\hbox{\hskip1pt$\sim$}}
\raise1pt\hbox{$<$}}}
\def\gsim{\mathrel{\rlap{\lower4pt\hbox{\hskip1pt$\sim$}}
\raise1pt\hbox{$>$}}} \def\sqr#1#2{{\vcenter{\vbox{\hrule height.#2pt
\hbox{\vrule width.#2pt height#1pt \kern#1pt \vrule width.#2pt} \hrule
height.#2pt}}}}
\def\square{\mathchoice\sqr66\sqr66\sqr{2.1}3\sqr{1.5}3}
\def\beq{\begin{equation}} \def\eeq{\end{equation}}
\def\beqa{\begin{eqnarray}} \def\eeqa{\end{eqnarray}}
\def\eq#1{Eq. (\ref{#1})}

\begin{document}
\title{Dark matter as a dynamic effect due to a non-minimal gravitational coupling with matter}

\author{O  Bertolami\footnote{Also at Departamento de F\'isica, Instituto Superior T\'ecnico.}, J. P\'aramos} 

\address{Instituto de Plasmas e Fus\~ao Nuclear, Instituto Superior T\'ecnico, \\ Avenida Rovisco Pais 1, 1049-001 Lisboa, Portugal}

\ead{orfeu@cosmos.ist.utl.pt, paramos@ist.edu}

\begin{abstract}
In this work the phenomenology of models possessing a non-minimal coupling between matter and 
geometry is discussed, with a particular focus on the possibility of describing the flattening of the galactic rotation curves as a dynamically generated effect derived from this modification to General Relativity. Two possibilities are discussed: firstly, that the observed discrepancy between the measured rotation velocity and the classical prediction is due to a deviation from geodesic motion, due to a non-(covariant) conservation of the energy-momentum tensor; secondly, that even if the principle of energy conservation holds, the dynamical effects arising due to the non-trivial terms in the Einstein equations of motion can give rise to an extra density contribution that may be interpreted as dark matter. The mechanism of the latter alternative is detailed, and a numerical session ascertaining the order of magnitude of the relevant parameters is undertaken, with possible cosmological implications discussed.
 
\end{abstract}

\section{Introduction}\label{intro}

Many modifications of the theories of gravity are strongly motivated by the issue of an alternative solution to the problem of the flattening of the galaxy rotation curves --- one which does not require dark matter in the strictest sense, {\it {\it i.e.}} weakly or non-interacting matter fields \cite{dm1,dm2,dm3,dm4}. With this goal in mind, standard $f(R)$ models \cite{fR1,fR2,fR3} have been studied endowed with power-law curvature terms $f_1(R) \propto R^n$ in the action, instead of the linear curvature depicted in the General Relativity (GR) action. For $n>1$, an addition to the Newtonian potential $\Delta \Phi(r) = -GM(r/r_c)^\be / 2r$ is found, with $\be$ a function of $n$ and $r_c$ being a parameter characteristic of 
each galaxy. A fit of several rotation curves indicates that $ n = 3.5$ (corresponding to $\be = 0.817$) yields the best agreement with observations \cite{CapoLSB}.

In another approach, it has been shown that an asymptotically flat velocity dispersion curve can be obtained if the curvature term in the Einstein-Hilbert action includes a logarithmic factor, $f_1(R) = f_0 R(1 + v \log R) $ \cite{LoboLog}; given the non-relativistic condition $v \ll 1$, this may be approached by a power-law with an exponent very close to unity, $f_1(R) \approx f_0 R^{1 + v^2}$.

Since this description does not take into consideration any galaxy-dependent parameters, an universal asymptotic velocity $v$ is obtained; this is in stark contrast with the Tully-Fisher and Faber-Jackson relations, which posit a dependence of the visible mass of a galaxy $M \propto v^m$, for spiral ($m=4$) and elliptical galaxies ($m=6$), respectively.

This work reviews the results obtained in Ref. \cite{paramos}, to which the reader is directed for a more technical discussion and the full set of computations. It comprises a small introduction to the non-minimal coupling model under scrutiny, proceeds with the implied non-geodesic motion already hinted above, and then proposes the already mentioned mechanism for dynamically mimicking dark matter; a numerical session and discussion follows.

\section{The model}
\label{model}

The non-minimal coupling between matter and geometry is embodied in the following action functional \cite{f2model},

\beq S = \int \left[ {1 \over 2}f_1(R) + [1 + \la f_2(R) ] \mathcal{L}_m \right] \sqrt{-g} d^4 x ,
\label{action} \eeq

\noindent where $f_i(R)$ (with $i=1,2$) are arbitrary functions of the scalar curvature $R$, $\mathcal{L}_m$ 
is the Lagrangian density of matter and $g$ is the metric determinant. The non-minimal coupling $f_2$ is gauged via the coupling constant $\la$ (with dimensions $[\la] = [f_2]^{-1}$). The standard Einstein-Hilbert action is recovered by taking $f_2=0$ and $f_1= 2 \ka (R - 2 \La)$, where $\ka = c^4 /16 \pi G$ and $\La$ is the cosmological constant (from now on, one works in a unit system where $c= 1$).

The modified Einstein equations of motion arise from variation of the action with respect to the metric $g_\mn$,

\beq \label{EE0} \left( F_1 + 2 \la F_2 \mathcal{L}_m \right) R_\mn - {1 \over 2} f_1 g_\mn =  
\left( \square_\mn - g_\mn \square \right) \left(F_1 + 2 \la F_2 \mathcal{L}_m \right) + \left( 1 + \la f_2 \right) T_\mn, \eeq

\noindent defining $\square_\mn \equiv \nabla_\mu \nabla_\nu$ for convenience, and $F_i(R) \equiv f_i'(R)$. The matter energy-momentum tensor is defined as

\beq \label{defTmn} T_\mn = -{2 \over \sqrt{-g}} {\de \left(\sqrt{-g} \mathcal{L}_m \right) \over \de g^\mn } ,
\eeq

\noindent so that the trace of \eq{EE0} reads

\beq \label{trace0} \left( F_1 + 2 \la F_2 \mathcal{L}_m \right) R - 2 f_1 = \\  -3 
\square \left(F_1 + 2 \la F_2 \mathcal{L}_m \right) + \left( 1 + \la f_2 \right) T  .\eeq

The Bianchi identities, $\nabla^\mu G_\mn = 0$ imply the covariant ``non-conservation law''

\beq \nabla^\mu T_\mn = {\la F_2 \over 1+ \la f_2} \left( g_\mn \mathcal{L}_m - T_\mn \right) 
\nabla^\mu R 
, \label{non-cons} \eeq

\noindent interpreted as a signal of energy-momentum exchange between matter and geometry due to the 
non-trivial terms \cite{scalar}.

For simplicity, and since one aims studying the phenomenology of the non-minimal coupling $f_2(R)$, one sets $f_1(R) = 2 \ka R $ (the GR form discarding the cosmological constant $
\Lambda$); \eq{EE0} then reduces to

\beq \label{EE}  \left( 1+ {\la \over \ka} F_2 \mathcal{L}_m \right) R_\mn - {1 \over 2} R g_\mn 
=   {\la \over \ka} \left( \square_\mn - g_\mn \square \right) \left( F_2 \mathcal{L}_m 
\right) + {1 \over 2 \ka}  \left( 1 + \la f_2 \right)  T_\mn, \eeq

\noindent and, equivalently, \eq{trace0} becomes

\beq \label{trace}  \left( 1 - {\la \over \ka} F_2 \mathcal{L}_m \right) R =  3{\la \over \ka} 
\square \left( F_2 \mathcal{L}_m \right) - { 1 \over 2 \ka}  \left( 1 + \la f_2 \right) T  .\eeq

\section{Flattening of the galaxy rotation curves due to non-geodesical motion}
\label{geo}

From a fundamental standpoint, the most striking characteristic of the model here considered is the non-conservation of the energy-momentum tensor of matter; assuming that matter is described by a perfect fluid distribution with the usual energy-momentum tensor

\beq T_\mn =\left( \rho +p\right) U_{\mu }U_{\nu }+p g_{\mu \nu } , \eeq

\noindent where $\rho$ is the energy density, $p(\rho)$ is the pressure (given by a suitable 
equation of state) and $U_\mu$ is the four-velocity (satisfying the normalization condition $U_\mu U^\mu = 
-1$), \eq{non-cons} leads to an extra force imparted upon a test particle,

\beq \label{force0}
f^{\mu}={1 \over \rho +p} \Bigg[{\la F_2 \over 1+\la f_2}\left({\cal L}_m+p\right)\nabla_\nu
R+\nabla_\nu p \Bigg] h^\mn,
\eeq

\noindent with the projection operator $h_\mn = g_{\mn}-U_{\mu}U_{\nu}$, so that 
$h_{\mn}U^{\mu }=0$ and the extra force is orthogonal to the four-velocity.

In GR, the specific form of the Lagrangian density does not appear explicitly in the equations of motion, but only via the derived energy-momentum tensor \eq{defTmn}. For this reason, and particularly if one is not interested in describing the thermodynamics of the perfect fluid, this functional is usually ignored. This, of course, changes in the present situation, since ${\cal L}_m $ appears explicitly in Eqs. (\ref{EE}) and (\ref{force0}).

In GR, the Lagrangian density ${\cal L}_m$ of a perfect fluid several equivalent on-shell 
expressions are admitted: for instance, ${\cal L}_{m0} = -\rho$, ${\cal L}_{m1} = p$ and ${\cal L}_{m2} = -na$, where $n$ is the particle number density, and $a$ is the physical free energy, defined as $a=\rho/n-Ts$, with $T$ the fluid temperature and $s$ the entropy per particle. However, in the presence of a non-minimal gravitational coupling with matter, this degeneracy is lifted and the original, bare Lagrangian density ${\cal L}_m = -\rho$ should be considered \cite{fluid,fluidfaraoni}.

One may further simplify computations by noticing that as matter in galaxies is non-relativistic (asides from the inner core region), and it may be described approximately by a pressureless dust distribution, $p \ll \rho c^2$. 

With the above consideration in mind, one now assumes a static, spherically or cylindrically symmetric {\it Ansatz} for the problem and that the velocity has only a tangential component, so that $h^{\mu r} = g^{\mu r} $ and so the extra-force \eq{force0} reads

\beq \label{force}
f^{\mu}= - {\la F_2 \over 1+\la f_2} R'(r) h^{\mu r },\eeq

\noindent where the prime denotes differentiation with respect to the radial coordinate $r$.

Endowed with the expression above, one may now t le the issue of describing the flattening of the galaxy rotation curves; instead of an asymptotic velocity dispersion $v(r) \rightarrow v_\infty$, one assumes that the reported flattening allows for a shallow decrease, {\it {\it i.e.}}, a profile of the form

\beq  v^2(r) = v_\infty^2 \left( {r_* \over r}\right)^\be, \eeq

\noindent where $r_*$ is an unknown normalization constant and $0 \leq \be < 1$ is the outer slope of the rotation curve. Identifying the extra-force with the observed deviation leads to the expression

\beq f^r \simeq - \la F_2 R'(r) -  {v_\infty^2 \over r} \left( {r_* \over r}\right)^\be, \label{problem} \eeq

\noindent which one may use to ascertain the form of the non-minimal coupling $f_2(R)$; in order to do so, one requires some bearing on the dependence of the scalar curvature on the radial distance $R= R(r)$; to do so, one recalls that it is implicitly assumed that the effect of the non-trivial terms appearing in \eq{EE} is perturbative --- so that the geodesics derived from the metric are similar to those found in GR --- and the non-minimal coupling manifests itself simply through an extra-force deviating test particles from geodesic motion.

As will be seen below, this approach is self-consistent; for now, one assumes that the scalar curvature is given by its classical expression $R = -8 \pi G T$, with $T = -\rho$ the trace of the energy-momentum tensor $T_\mn$. Thus, the issue of obtaining a dependence $R(r)$ simplifies to ascribing a particular profile for the density distribution $\rho(r)$ of matter. For this, one takes a power-law profile of the form

 \beq \rho(r) \approx \rho_0 (a/ r)^m,\eeq 
 
 \noindent with $m\neq 0$ and $\rho_0$, $a$ characteristic scales.
 
With the differential equation \eq{problem} in a closed form, one may solve for $f_2(r)$ and, reversing the dependences above, find the aimed coupling $f_2(R)$. After some algebraic steps (detailed in Ref. \cite{paramos}), one finds that 

\beq f_2(R) =
\cases{-{1 \over \be} \left({R \over R_*} \right)^{\be/m} &, $\be \neq 0$
\cr -{1 \over m} \log \left( {R \over R_*} \right)  &, $\be = 0$
},
\label{logR}
\eeq

\noindent defining the characteristic scale $ R_* = 8 \pi G \rho_0(a/r_*)^m $.

With the above solution, it is trivial to show that the conditions for a negligible effect of the non-minimal coupling $f_2(R)$ on the Einstein field \eq{EE},

\beq |\la f_2(R)| \ll 1 ~~,\qquad  \la \left[F_2(R) \rho \right]'' \ll \rho, \label{pert} \eeq

\noindent (using that $T = {\cal L}_m =-\rho$) are satisfied for a wide domain, $ 1 ~kpc \lesssim r \lesssim 50 ~kpc$, where the flattening of the rotation curves is measured for a wide range of values of $r_*$ (mainly due to the smallness of the gauging constant $\la = v^2_\infty \approx 10^{-12}$): one could set, e.g., $r_*= 1~kpc$.

As a result, the assumption of a perturbative regime for the equations of motion \eq{EE} is indeed self-consistent --- although the 
deviation from geodesical motion is relevant, dominating the Newtonian acceleration and leading to the flattening of the galaxy rotation curve.

Before continuing, one remarks that a similar exercise would ensue if one was to assume that the velocity dispersion curve follows a logarithmically decaying profile,

\beq v^2(r) = v_\infty^2  {r_* \over r} \log \left( {r \over r_*} \right),\eeq

\noindent instead of the simpler power-law assumed above. Indeed, and in the context of standard galactic dark matter descriptions, this radial dependence leads to a dark matter density distribution given by the so-called Navarro-Frenk-White (NFW) density profile, which is widely used in the literature \cite{NFW} (see Appendix A of Ref. \cite{paramos} for a discussion of different density profiles). Assuming this decay law for $v^2(r)$ leads to the non-minimal coupling 

\beq f_2(R) = - \left({R \over R_*}\right)^{1/m} \left[1 + {1 \over m} \log \left({R_* \over R}\right) \right]. \eeq

\noindent and, as before, the perturbative conditions \eq{pert} are trivially satisfied within the relevant range.

To close this section, one first remarks that  the obtained expression \eq{logR} for the non-minimal coupling $f_2(R)$ in the flattened rotation curve case $\be = 0$ is strikingly similar to that of Ref. \cite{LoboLog}, where the non-trivial curvature term is given by $f_1(R) = R[1+v^2 \log(R/R_*)] \approx R (R/R_*)^{v_\infty^2}$. Following that proposal, one may suggestively rewrite the action \eq{action} as

\beq S = \int \left[ 8\pi G R + \left( {R_* \over R } \right)^\alpha \mathcal{L}_m \right] \sqrt{-
g} d^4 x ,
\label{action2} \eeq

\noindent defining $\al = v_\infty^2 / m$.

However, this similarity with Ref. \cite{LoboLog} leads to the same caveat, namely an undesired universality of the asymptotic value of the predicted galaxy rotation curves: assuming similar outer slopes $m$ for the visible matter density profile $\rho$ of galaxies, one concludes that $v_\infty$ and $\be$ should be fixed parameters of the model, unadjustable for different galaxies. This is not observed for the available rotation curves, in which the asymptotic value $v_\infty$ may vary somewhat (although of the order of magnitude $v_\infty \sim 100 km/s$). One may assume that for different density profiles, spatial symmetry and morphology considerations would contribute to alleviate this issue, but the obtained universality remains a noticeable disadvantage of this approach --- thus prompting the search for alternatives within the ample phenomenological arsenal provided by the model under scrutiny.

\section{Pressureless dust with non-minimal gravitational coupling}
\label{metric}

Motivated by the considerations put forward in the last paragraph, one now assumes the opposite stance: instead of considering a perturbative effect of the non-minimal coupling, but a large deviation from geodesic motion, one posits that the energy-momentum tensor of matter is approximately conserved, but the geodesics thus followed by test particles differ significantly from those of GR. As in the previous section, this approach also proves to be self- consistent.

Given the above, the problem becomes similar to those faced by standard $f(R)$ theories (that is, with $f_2(R)=1$): solve the Einstein field equations so that the obtained metric gives rise to the observed galaxy rotation curves. Since one ultimately aims to describe dark matter as a dynamical effect stemming from the action functional \eq{action}, it is productive to try to express the effect of the non-minimal coupling in terms of an equivalent, mimicked dark matter density. This also allows for a direct comparison with existing profiles for the distribution of dark matter, arising both from simulations as well as observational evidence.

For generality, a power law coupling of the form

\beq f_2(R) = \left( {R \over R_0} \right)^n, \label{f2}, \eeq

\noindent is assumed, with $n$ a yet unspecified exponent and $R_0$ a characteristic scale (which amounts to a rewriting of the gauging constant $\la$).

It should be highlighted that \eq{f2} does not exclude a more elaborate form for the non-minimal coupling, which could then be written as a series

\beq f_2(R) = \sum_p \left({R\over R_{0p}}\right)^p ,\eeq

\noindent summed over the rational exponents $p$ of the ``full'' model (integer or not); if this is the case, the choice 
of \eq{f2} is to be viewed as a valid approximation within the galactic context (that is, the dominant term of the series for the corresponding range of $R$). For this reason, this assumption is not in contradiction with the previous work by the authors \cite{matter}, which approached the possibility of including a 
linear coupling $f_2 = R/R_1$ (assumed relevant in a high density environment such as the Sun).

Again, one models matter within a galaxy as a pressureless perfect fluid,  {\it i.e.}, dust,
characterized by a density $\rho$; the corresponding energy-momentum tensor thus reads

\beq T_{\mu \nu } = \rho U_{\mu }U_{\nu } \rightarrow T = -\rho. \eeq

\noindent Using the Lagrangian density ${\cal L}_m =  - \rho$ \cite{fluid}, \eq{EE} reads

\beq \label{EE1} \left[ 1 - {n \over \ka} \left({R \over R_0} \right)^n {\rho \over R} \right] R_\mn 
- {1 \over 
2} R g_\mn =  {n \over \ka} \left(  g_\mn \square -\square_\mn \right) \left[ \left({R 
\over 
R_0} \right)^n {\rho \over R} \right] + {1 \over 2 \ka}  \left[ 1 + \left({R \over R_0} 
\right)^n 
\right] \rho U_\mu U_\nu, \eeq

\noindent and \eq{trace} becomes

\beq \label{trace1} R = {1 \over 2 \ka} \left[ 1 + (1 - 2n) \left({R \over R_0} \right)^n \right] \rho 
-  {3n \over \ka} \square \left[ \left({R \over R_0} \right)^n {\rho \over R} \right]  , 
\eeq

As stated at the beginning of this section, the additional terms induced by the non-minimal gravitational coupling $f_2(R)$ are interpreted 
as contributing to the dark matter halo density profile $\rho_{DM}$. This can be better ascertained by ignoring the matter contribution arising from GR, so that the non-minimal terms dominate far away the galactic core (where the rotation curve flattens). In this exterior region, one may rewrite \eq{trace1} as

\beq \label{trace2} R \simeq  {1 -2n\over 2 \ka} \left({R \over R_0} \right)^n \rho - 3 {n\over \ka} \square 
\left[ \left({R \over R_0} \right)^n {\rho \over R} \right]  . \eeq

\subsection{Static solution}

Assuming a known visible matter density $\rho$, solving the above equation amounts to 
obtaining the relation $R = R(\rho)$; this is somewhat analog to the previous exercise, where one assumed $\rho(r)$ and computed $R(r)$. Naturally, the classical identification $R = 2 \ka \rho$ fails, since it would lead to a differential equation for the visible matter density which might not be satisfied by the considered 
profile $\rho(r)$. Inspection of \eq{trace2} shows that a solution defined implicitly by the expression

\beq R = {1 -2n\over 2 \ka} \left({R \over R_0} \right)^n \rho ,\eeq

\noindent is self-consistent, since the gradient term vanishes everywhere; accordingly, this is dubbed as a ``static'' solution. This yields

\beq \label{solution} R = R_0 {\left[ {1 - 2n \over 2 \ka} {\rho \over R_0}\right]}^{1/(1-n)}= R_0 
{\left[ (1 - 2n ) 
{\rho \over \rho_0}\right]}^{1/(1-n)} ,\eeq

\noindent introducing the characteristic density $\rho_0 \equiv R_0/2\ka$.

In order to interpret the additional terms shown in the Einstein field Eqs. (\ref{EE1}), one must first show that the above ``static'' solution provides a suitable description of dark matter, {\it i.e.} the full tensorial aspect of the problem must be taken into account, not merely the algebraic identification stemming from the trace \eq{trace2}. Indeed, replacing \eq{solution} in \eq{EE1} allows one to conclude that the dynamic effect due to the non-minimal coupling behaves as a perfect fluid characterized by an energy-momentum tensor analog to that of matter, so that this mimicked ``dark matter'' shares the same 4-velocity as visible matter, but has a pressure and density given by

\beq \label{rhodm} \rho_{dm} = {1-n \over 1-4n} \rho_0 {\left[ (1 - 2n ) {\rho \over 
\rho_0}\right]}^{1/(1-n)} , 
\eeq

\noindent thus yielding a non-vanishing pressure and an equation of state of the form

\beq \label{EOS} p_{dm} = {n \over 1-n} \rho_{dm}. \eeq

Hence, at a sufficiently large distance from the galactic core, where the rotation curve 
has flattened, the non-minimal gravitational coupling with ordinary matter mimics a dark 
matter component which is dragged by ordinary matter, and may have a negative density $\rho_{dm} < 0 $, if $1/4 < n 
< 1$; although strange, this is not a pathology of the model, as the curvature in the considered outer 
region will remain positive, $R > 0$. As will be seen below, relevant models for the density profile of dark matter lead to negative exponents $n$, so that the density is indeed positive defined.

\subsection{Mimicking dark matter density profiles}
\label{profiles}

Many dark matter models resort to the assumption that, at the region where the galaxy rotation curve has flattened, both visible and 
dark matter might be modelled by a power law density profile, characterized by the outer slopes $m$ 
and $m'$, respectively (see Appendix A of Ref. \cite{paramos} for a thorough discussion),

\beq \label{approx} \rho \approx \rho_v \left({a \over r}\right)^m ~~,\qquad \rho_{dm} \approx 
\rho_d \left({a' \over r } \right)^{m'}. \eeq

Thus, one may establish the algebraic relation $\rho_{dm} 
\propto \rho^p $, defining the slope ratio $p \equiv m' / m$. The constant $\rho_v$ is specific to each visible matter distribution $\rho$, 
and $\rho_d$ will be related to this quantity and to the previously defined $\rho_0$. Since a similar power law was obtained in the previous 
paragraph, this relation allows for a direct translation between the parameters $R_0$, $n$ of the non-minimal coupling model and the mimicked dark matter density profile quantities $a'$, $m'$ and $\rho_d$, assuming a given visible matter density profile parameters, 

\beq \label{exprel} {1 \over 1-n} = {m' \over m} \equiv p \rightarrow n = 1 - {m \over m'} = 1- 
{1\over p},
\eeq

\noindent  and

\beq \rho_d = {1 \over 4+3p} \left({2 \over p}-1 \right)^p \rho_0^{1-p} \rho_v^p. \eeq

\noindent One may assume that the length scale $a'$ is inherited from the visible matter 
density, so that $a' \sim a$.

Clearly, the long range dominance of the mimicked dark matter over visible matter indicates that the density profile of the former should be shallower than that of the former, so that $m' < m$. Hence, one posits that $ p \equiv m'/m < 1$, thus a negative exponent $n$ is expected. This range is of physical significance: since dark matter is relevant at large distances, but the scalar curvature increases towards the galactic core, the non-minimal coupling $f_2(R)$ should scale with an inverse power of the curvature.
 
For concreteness, one ascertains what is the relevant range for the exponent $n$; to do so, one has to assume a particular visible matter density profile. In this study, the Hernquist distribution is considered \cite{Hernquist},

\beq \label{Hernquistpro} \rho(r) = {M \over 2\pi }{a \over r} {1 \over \left(r+a \right)^3} , \eeq

\noindent so that $m = 4$, and \eq{exprel} leads to $n = 1 - 4/m'$. Bearing in mind that one aims replicating the dark matter distribution, it is useful to resort to two widely used candidates, namely the NFW profile

\beq \rho_{dm}(r) =  {\rho_{cp} \over \left({r \over a} \right) \left(1 + {r \over a} \right)^2 }  \eeq

\noindent with $m' = 3 \rightarrow n = -1/3 $, and the isothermal sphere profile 

\beq \rho(r) = {v_\infty^2 \over 2\pi G r^2},\eeq

\noindent with $m' = 2 \rightarrow n = -1$.

Both dark matter models are characterized by negative exponents $n$, as indicated above. The obtained values $n = -1/3$ and $n= -1$ should be taken into consideration in the following developments, although the general exponent $n$ is retained.

\subsection{Background matching}
\label{matching}

The elegant ``static'' solution obtained before can be fruitfully used to determine several characteristics of the model, as has been extensively discussed in Ref. \cite{paramos}; most calculations are hence omitted, with the focus shifted to a comprehensive discussion of the physical significance of these results.

Firstly, one notices that a negative power-law coupling $f_2= (R/R_0)^n$ (with $n<0$) does not admit an asymptotically flat Minkowski space-time, since it would lead to a singular $R=0$ scenario. This problem is naturally avoided by noticing that the assumed coupling is valid at the galactic scale, and that the asymptotic behaviour should be identified with the large scale, cosmological background. Hence, the notion of a matching distance $r_\infty$ natural appears, defined so that the density of the mimicked dark matter approaches the order of the critical density, $\rho_{dm} (r_\infty) \sim \rho_c$ (the latter being is defined as $\rho_\infty= 3H_0^2/(8\pi G)$, with $H_0 \simeq 70~(km/s)/Mpc$ the Hubble constant). Using \eq{solution} yields

\beq \label{matchingr}  r_\infty = \left[{2\over 3} \times 3^n (1-2n) 
\right]^{1/4} \left( {c \over r_0 H } \right)^{(1-n)/2} \left( r_s r_0^2  a\right)^{1/4} ,  \eeq

\noindent having defined the characteristic lengthscale $r_0 = 1/\sqrt{R_0}$. The obtained matching distance which may be regarded as an indicator of the radius of the mimicked ``dark'' matter halo; for consistency, it should vanish when the non-minimal coupling \eq{f2} is negligible,  {\it i.e.} $R_0 \rightarrow 0$ if the exponent $n$ is negative or $R_0 \rightarrow \infty$ for $n>0$; this is trivially satisfied, since $r_\infty \propto r_0^{n/2} = R_0^{-n} \rightarrow 0$ (excluding the 
constant coupling $n=0$ case).

One may also define the characteristic density 

\beq \label{rhoinfty} \rho_0 = 2 \ka R_0 = {c^2 \over 8 \pi G r_0^2 } = \left({c \over r_0 H }\right)^2 {\rho_
\infty \over 3} = \left({2.5 ~Gpc \over r_0 }\right)^2 \rho_\infty . \eeq

\noindent Notice that the equation of state previously found \eq{EOS} enables a negative pressure for any negative exponent $n$; furthermore, in a cosmological setting, the non-trivial coupling $f_2(R) = (R/R_0)^n $ is given by $( \rho_\infty / \rho_0)^n $. Hence, the obtained relation \eq{rhoinfty} does hint a putative unified description of dark matter and dark energy, responsible for the acceleration of the expansion rate of the Universe. Other models of unification of dark energy and dark matter include the Chaplygin gas model \cite
{Chaplygin1, Chaplygin2, Chaplygin3, Chaplygin4} and some field theory constructions \cite{Rosenfeld}.

\subsection{Crossover between visible and dark matter dominance}
\label{crossover}

Proceeding with the analysis of the obtained ``static'' solution, one remarks that the model parameter $R_0$ introduces both the 
length scale $r_0 $ as well as a density scale $\rho_0$; the former should have a clear physical meaning. Indeed, $r_0$ is associated with the crossover distance $r_c$ at which the density of the mimicked dark matter component becomes similar to that of visible matter, thus signalling the onset of the flattening of the galaxy rotation curves. By identifying $\rho_{dm}(r_c)  = \rho(r_c)$ and using the obtained profiles \ref{rhodm} and \eq{Hernquistpro}, one obtains 

\beq  \label{rc} r_c^4  = 2 (1-2n)^{1/n} r_s r_0^2 a, \label{r_c} \eeq

\noindent with $r_s = 2GM/c^2$ being the Schwarzschild radius of the galaxy. Hence, the crossover radius $r_c$ is proportional to a sort of ``geometric mean'' of all the length scales present.

\subsection{``Gradient'' solution}
\label{gradient}

One ends this section by discussing the relevance of the gradient term on the {\it r.h.s} of \eq{trace1}; indeed, although the ``static'' solution makes it vanish identically, it was obtained by neglecting the contribution from visible matter, since one is assuming that dark matter is dominant at large distances. Conversely, at smaller distances one expects the solution to \eq{trace1} to deviate significantly from the ``static'' solution \eq{solution}, interpolating between the latter and the classical identification $R = 8 \pi G \rho$. Thus, in the regime before the onset of dark matter dominance $r < r_c$, the contribution from the mentioned gradient terms may be highly relevant.

This may be justified by inserting the GR identification $R = 8 \pi G \rho$ into \eq{trace1}, and reading the gradient terms. 
To do so, one assumes that the D'Alembertian operator is given by its expression in a flat spacetime (in spherical coordinates), which amounts to neglecting relativistic corrections. This requirement translates into the condition  $r_s < r_0$, which is easily satisfied by the mechanism under scrutiny, as shall be seen.

A somewhat evolved analysis depicted in Ref. \cite{paramos} indeed shows that the gradient term may become dominant below a distance $r_k < a < r_c$, 

\beq r_k \equiv \left[ 6n^2 (1-n) \left( {a^2 \over 2 r_s} \right)^{1-n} r_0^{2n} \right]^{1/(n+1)}. \eeq

\noindent Notice that, at even shorter distances, relativistic corrections of the form $r_s/r$ should be considered. It can be shown that this is not related to the particular form of the visible matter density distribution (in this case, the Hernquist profile), and that assuming another form for $\rho(r)$ would still lead to an eventual dominance of the gradient term of \eq{trace1} below a threshold $r_k$.

This concern may be attenuated by noticing that one is assuming the validity of the Newtonian regime: if the lengthscales $a$, $r_s$ and $r_0$ yield a value for $r_k$ smaller than the Schwarzschild radius, then the need to include relativistic corrections to the computation of the gradient term makes the whole approach inconsistent. As a numerical session shall show, the condition $r_k > r_s$ is not always satisfied, so that the above expression for $r_k$ should be considered with due care.

Skipping over the technicalities, it may be shown that it is impossible to simultaneously ensure that the gradient term on the {\it r.h.s} of \eq{trace1} is negligible for all distances: it dominates at either $r<a$ or $r > a$. This is not unexpected, given that the lengthscale $a$ signals the shift of behaviour of the Hernquist profile \eq{Hernquistpro}. Hence, one is forced to conclude that the ``static'' solution \eq{solution} is not valid after all! In particular, the discussion of the dominance of the gradient term in the outer region $a < r < r_c$ is flawed, as the crossover distance $r_c$ assumed for its validity.

If the gradient term indeed dominates the {\it r.h.s} of \eq{trace1} at short ranges, it turns out that the corresponding ``gradient'' solution to this equation is the relevant quantity to assess, and not the ``static'' one already described. By the same token, it is legitimate to assume that this ``gradient'' solution will play the role of dark matter, so that it should dominate the contribution of visible matter at long ranges. Thus, \eq{trace1} may be approximated by

\beq \label{tracekin} R \approx-  {3n \over \ka} \left( {d^2 \over dr^2} + {2 \over r} {d \over dr} \right) \left[ \left({R \over R_0} \right)^n {\rho \over R} \right] .
\eeq

One now introduces the redefined field

\beq \label{redef} \varrho =  \left( {R \over R_0}\right)^{-n/(1-n)}, \eeq

\noindent  and coordinate $y = r_0 /r$; the dominance of this ``gradient'' solution over the visible matter contribution occurs in the range $r > r_k $, that is, $ y < y_k$. 

In order to better understand the evolution of this ``gradient'' solution, one may compute the order of magnitude of the quantities involved at $y = y_k$ (by recalling that, for smaller physical distances $y > y_k$, the classical solution $R = 8 \pi G \rho$ is valid and inserting the Hernquist profile for $\rho(r)$). One finds that 

\beq \label{tracekin2}  \left( {d^2 \varrho^{n-1} \over d y^2 }\right)_{y=y_k} = -{1 \over 3n} \left({r_s a \over r_0^2 }\right).\eeq

\noindent Inserting typical values for galaxies $a=(1-10)~kpc$ and $r_s =  (10^{-3}-10^{-1})~pc$ (corresponding to galaxies of approximately 
$(10^{10}-10^{12})$ solar masses) and recalling the validity condition for the Newtonian regime, $r_0 > r_s$, one concludes that the {\it r.h.s.} of the above equation is less than unity; it turns out that the values involved are extremely small (of the order $10^{-14}$ or lower), so that one may safely consider that \eq{tracekin} is essentially given by 

\beq {d^2 \varrho^{n-1} \over d y^2 }\approx 0 \eeq

\noindent at large distances, {\it i.e.} small values of $y$ --- the full form of \eq{tracekin} with the above redefinitions would be that of a generalized Emden-Fowler differential equation. Hence, the ``gradient'' solution is very well approximated by $\varrho = {\rm const.}$; recalling \eq{redef} leads to 

\beq R \propto R_0 \left( \rho \over \rho_0\right)^{1/(1-n)}, \eeq

\noindent that is, the ``gradient'' solution tracks the ``static'' solution! Hence, the previous results are partially vindicated, and the identification of the induced effect due to the non-minimal coupling $f_2(R)$ with a perfect fluid with equation of state \eq{EOS} is still valid.

\subsection{Tully-Fisher law}

The Tully-Fisher law is an empirical relationship between the intrinsic luminosity $L$ and a power of the rotation velocity 
$v_\infty$ of a spiral galaxy. An analogous relationship exists for elliptical galaxies, the Faber-Jackson relation. Since the luminosity is proportional to 
the visible mass, both may be written as $M \propto v_\infty^\sigma$, with $3 < \sigma \lesssim 4$. 

If, for simplicity, one assumes that $M_{dm}(r) \propto r$ at large distances, thus producing a flat rotation curve, the familiar $v_\infty^2 \propto M_{dm}$ relation 
is recovered; resorting to the scaling \eq{rhodm}, this yields

\beq \label{TF} M \propto v^{2(1-n)} , \eeq

\noindent so that the exponent of the Tully-Fisher or Faber-Jackson relations is related to that of the non-minimal coupling $f_2(R)$. For the scenarios already discussed, one obtains $M \propto v^{8/3}$, for $n=-1/3$, mimicking the NFW dark matter density profile, and $M \propto v^4$, for $n=-1$, mimicking a dark matter isothermal sphere.

\subsection{Conservation of energy-momentum tensor}

Section \ref{geo} depicted a first attempt to describe the flattening of the galactic rotation curves as due to the non-conservation of 
the energy-momentum tensor of matter, {\it i.e.}, through the non-geodesical motion experienced by fluid particles. Conversely, one now explores the possibility of mimicking ``dark matter'' as a consequence of the dynamics of \eq{trace1}. As explicitly stated in the beginning of section \ref{metric}, one is assuming that matter follows geodesics, so that the solution of \eq{trace1} gives rise to the observed rotation curves.

This assertion may now be verified, by rewriting \eq{force0}, with $p=0$ for pressureless matter, the Lagrangian density ${\cal L}=-\rho$ and 
the adopted non-minimal coupling $\la f_2(R) = (R/R_0)^n$:

\beq \label{force01}
f^{\mu}=  -n { \left( {R \over R_0}\right)^n \over 1+\left( {R \over R_0}\right)^n} \left[ \log\left({R \over R_0} \right)\right]_{,\nu} h^
\mn,
\eeq

\noindent with radial component 

\beq \label{force01r}
f^r =  -n { \left( {R \over R_0}\right)^n \over 1+\left( {R \over R_0}\right)^n} \left[ \log\left({R(r) \over R_0} \right)\right]',
\eeq

\noindent assuming spherical symmetry and the Newtonian approximation $h_{rr} = g_{rr} - (v_r/c)^2 \sim 1$.

As discussed in Ref. \cite{paramos}, a clear distinction between the ``static'' and the ``gradient'' solutions is that, although both are essentially the same (up to a multiplicative factor), each arises from widely different assumptions: the former assumes that the gradient terms on the {\it r.h.s.} of \eq{trace1} may be disregarded, so that the dominance of dark matter translates into the condition $(R/R_0)^n > 1/(1-2n)$; the latter, however, is based upon the dominance of these gradient terms, which implies the converse relation, $(R/R_0)^n < 1/(1-2n) < 1$, for negative exponent $n$.

Thus, one concludes that the issue of the covariant conservation of the energy-momentum tensor of matter branches between both solutions: from \eq{force01r}, one sees that the ``static'' solution would lead to 

\beq \label{force01rnot}
f^r=  -n \left[ \log\left({R \over R_0} \right)\right]' = -n {R'\over R} =  {n m' \over r}.
\eeq

\noindent since $R \propto \rho_{dm} \propto r^{m'}$ in the outer region of galaxies; given that $n$ is negative, this force correctly points towards the galactic center, thus enabling for a flattening of the rotation curve with a constant asymptotic velocity. However, for small $|n|$ and $m'$, this quantity would be given by $v_\infty /c = \sqrt{-nm'} \sim 1$, in blatant violation of the observed value $v_\infty \sim 10^2~km/s \sim 10^{-3}c$! Not only is this clearly unphysical (and would undermine the Newtonian approach taken throughout the text), but it also negates the whole purpose of this section, namely to assume geodesic motion.

Contrariwise, the verified dominance of the ``gradient'' solution and related condition  $(R/R_0)^n < 1/(1-2n) < 1$ leads to a radial force 

\beq \label{force01rdyn}
f^r =  -n \left({R \over R_0}\right)^{n-1} \left({R(r) \over R_0} \right)'.
\eeq

\noindent and the pathological result stemming from \eq{force01rnot} is avoided. 

For the already considered values of $a=(1-10)~kpc$ and $r_s =  (10^{-6}-10^{-4})~kpc$, it suffices that $r_0 > 200~kpc$ for the dominance of this extra force over the Newtonian one to occur only for distances greater than $1~Mpc$. Hence, this extra force  is negligible at galactic scale, and given the actual values for $r_0$ that will be used afterwards, much greater than this lower limit. Moreover, the energy-momentum tensor is conserved to a very good approximation.

\subsection{Energy conditions}
\label{energyconditions}

One now checks if the strong, null, weak and dominant energy conditions (SEC, NEC, WEC and DEC, respectively) are verified; following Ref. \cite{energy}, a little algebra yields the condition

\beq \left(1 + { E_n \over 1 + \left( {R \over R_0} \right)^{-n} } \right)\rho \geq 0 ,\eeq

\noindent with 

\beq E_n=
\cases{-2n &, {\rm SEC}
\cr 0 & , {\rm NEC}
\cr 2n & , {\rm DEC}
\cr n & , {\rm WEC}
}.
 \label{cases}
\eeq

Clearly, both the NEC and SEC are always satisfied, for a negative power law $n< 0$. The WEC and DEC are also fulfilled for the 
ranges $-1 \leq n \leq 0$ and $-1/2 \leq n \leq 0$, respectively. Since the  condition $R > R_0$ is always valid at galactic scales (due to the dominance of the gradient term of 
\eq{trace1}) for the considered models $ n= -1$ (isothermal sphere profile) and $n=-1/3$ (NFW profile), one has that $ \left( 1+ (R/R_0)^{-
n} \right)^{-1} \approx 0$; hence, the WEC and DEC are also satisfied for these scenarios.

A final check is in order, namely whether the considered mechanism for the mimicking of dark matter does not give rise to the so-called
Dolgov-Kawasaki instabilities, whereas small curvature perturbations could arise and expand uncontrollably \cite{DK}. Refs. \cite{energy,viability} allow one to write the mass scale $m_{DK}$ for these as

\beqa \label{mDK} m_{DK}^2 &=& {2\ka - 2 {\cal L}_{m} \left[ f'_2(R) + f_2''(R) R \right] + f_2'(R)T \over 2 {\cal L}_{m} f_2''(R) } = \\ 
\nonumber &=&- {2 \ka R \left( {R \over R_0} \right)^{-n} + n(2n-1) \rho \over 2n(n-1) \rho} R , \eeqa

\noindent which are given approximately by

\beq m^2_{DK}  \approx \left( {R \over R_0} \right)^{2-n} {\rho_0 \over 
\rho} { r_0^{-2} \over 2n(n-1) }  . \eeq

\noindent Since $m^2_{DK}$ is negative, the possibility of instabilities at a length scale $r_{DK}$ is worrisome, with

\beq r_{DK} = m_{DK}^{-1} =  \sqrt{2n(n-1)} \left( {R_0 \over R} \right)^{1-n/2} \sqrt{\rho \over \rho_0}  r_0 . \eeq

\noindent At large distances, where the dark matter-like component dominates and the scalar curvature is greater than the matter 
density, one finds that this typical length scale $r_{DK}$ grows linearly for the $m'=2$ isothermal sphere scenario, and with a power-law $r^
{3/2}$ for the $m'=3/2$, corresponding to the NFW density profile.

It should be noted that the generation of these instabilities assumes an initial small perturbation in a constant background curvature; this is best suited for an analysis in cosmology, but should be approached with due care in the present context, where there is a spatial variation of $R(r)$. Indeed, one must instead resort to the characteristic lengthscale over which one may assume a constant curvature, given by $L \equiv -R(r)/R'(r) = r/m' $. One may compare this constant curvature domain $L$ with the instability length scale $r_{DK}$ to ascertain if  the assumed perturbative expansion, as embodied in the inequality $r_{DK} < L$, is valid. Equating both quantities yields 

\beq \label{insta} \left({r\over a}\right)^{1+n\over1-n} < \varrho^{-n+1/2} (2\de \bar{a} )^{1\over 2(1-n)} {a \over 4 r_0 }\sqrt{n-1\over 
2n} . \eeq

\noindent This relationship immediately shows that the perturbative expansion is invalid if  $n=-1$, when dark matter assumes a isothermal sphere distribution; if it takes the form of a NFW profile, $n = -1/3$, one finds that the perturbative expansion is valid only below distances $r< r_s$, where relativistic corrections become relevant. Hence, the Dolgov-Kawasaki instabilities are not physically meaningful in the context of this work.

\section{Mimicking the dark matter density profile of spherical galaxies}

This section aims to apply the developed mechanism for the mimicking of dark matter to fit numerically several rotation curves of 
galaxies of E0 type --- these are chosen since they exhibit an approximate spherical symmetry adequate for the results obtained in this work. Seven 
galaxy rotation curves, are used, with the division into visible and dark matter components as reported in Refs. \cite{kronawitter, kronawitter2,Faber,Rix}: NGC 2434, NGC 5846, NGC 6703, NGC 7145, NGC 7192, NGC 7507 and NGC 7626 galaxies.

As stated in the preceding section, particular interest is assigned for $n = -1$ and $n = -1/3$, which correspond to the isothermal sphere or NFW dark matter density profiles. Furthermore, one may recall that the power-law profile $f_2 (R) = (R / R_0)^n$ was discussed in the context of a possible series expansion of a more general non-minimal coupling; hence, there is no reason to select {\it a priori} either case $n = -1$ or $n=-1/3$: instead, one can successfully take advantage of a composite coupling given by

\beq \label{f2two} f_2(R) =\sqrt[3]{R_3 \over R} + {R_1 \over R} , \eeq

\noindent so that both profiles may be added in order to describe the dark matter density profile and the related galaxy rotation curve.

The proposed non-minimal coupling leads to a more complex expression for the trace of the field equations

\beq \label{tracetwo} R = {1 \over 2 \ka} \left[1 + 3{R_1 \over R} + {5 \over 3} \left({R_3 \over R}\right)^{1/3} \right] \rho + {1 \over \ka} 
\square \left( \left[3 {R_1 \over R} + \left({R_3 \over R}\right)^{1/3} \right] {\rho \over R} \right)  .\eeq

\noindent One can no longer resort to the identification of a ``static'' solution, as before, since the non-linearity of 
the above equation indicates that the solution for this composite scenario is not given by the mere addition of individual solutions; by the same token, no clear equation of state or identification of the mimicked dark matter profile with a perfect fluid arises. Nevertheless, one expects that the main feature of the model are kept, namely the overall mimicking of dark matter density profiles according to \eq{rhodm} and the dominance of the gradient term on the {\it r.h.s.} of the above equation.

This stated, one proceeds as follows: each visible matter density profile $\rho(r)$ is derived from the visible component of the corresponding galaxy's rotation curve; two added Hernquist profiles are used, one for the extended gas and another for the core matter distribution. This density profile is then used as an input to \eq{tracetwo}, and a best fit scenario for the dark matter component of the rotation curve is obtained by varying the parameters $r_1 = 1/\sqrt{R_1}$ and $r_3 = 1/\sqrt{R_3}$. Although these should be universal quantities, one resorts to individual fits of each galaxy, allowing for a posterior discussion of order of magnitude, deviation between obtained values, any abnormal case or possible trends, {\it etc}.

\begin{figure}[h]
\begin{center}
$\begin{array}{cc}

	\includegraphics[width=7.5cm]{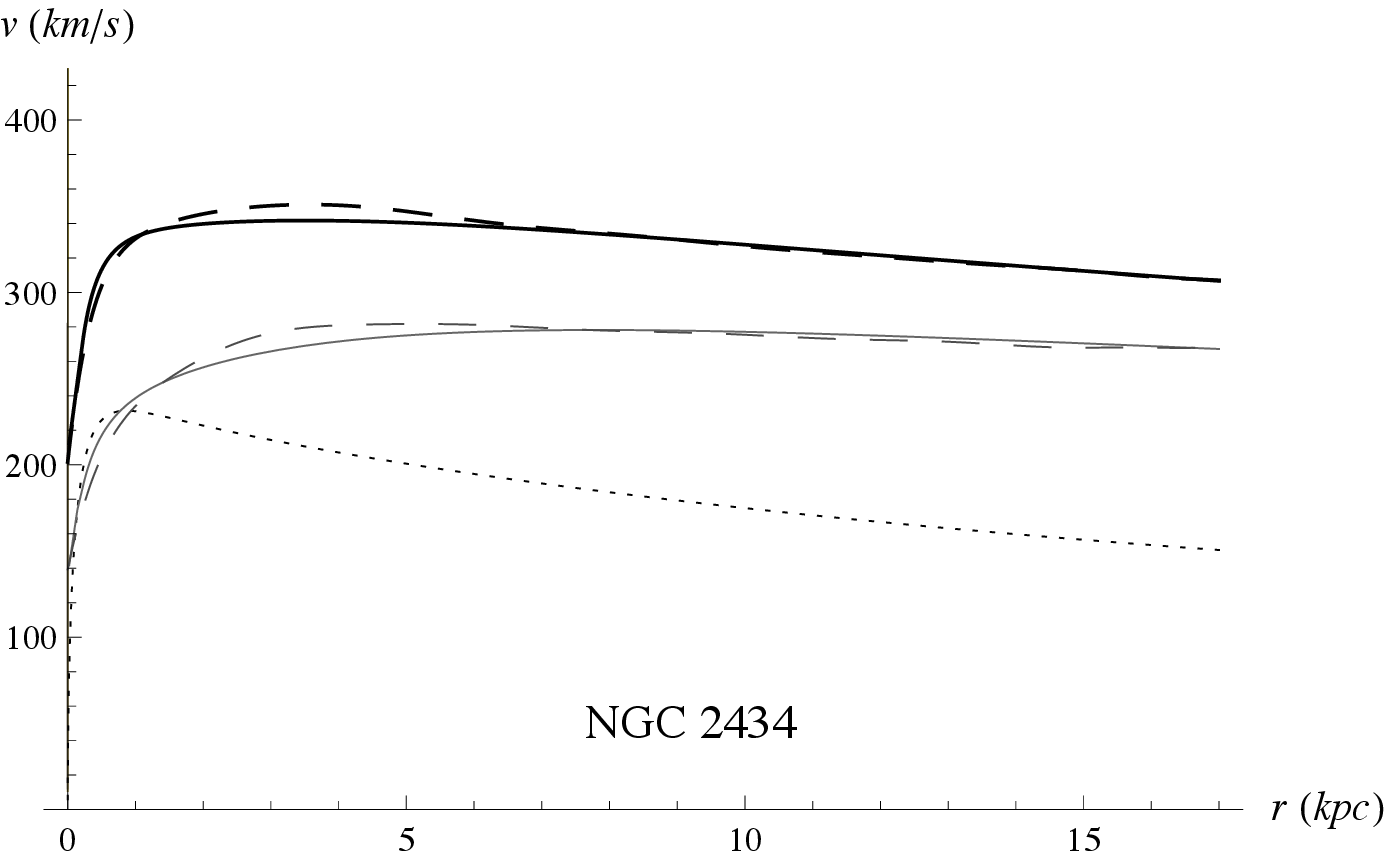} &
	\includegraphics[width=7.5cm]{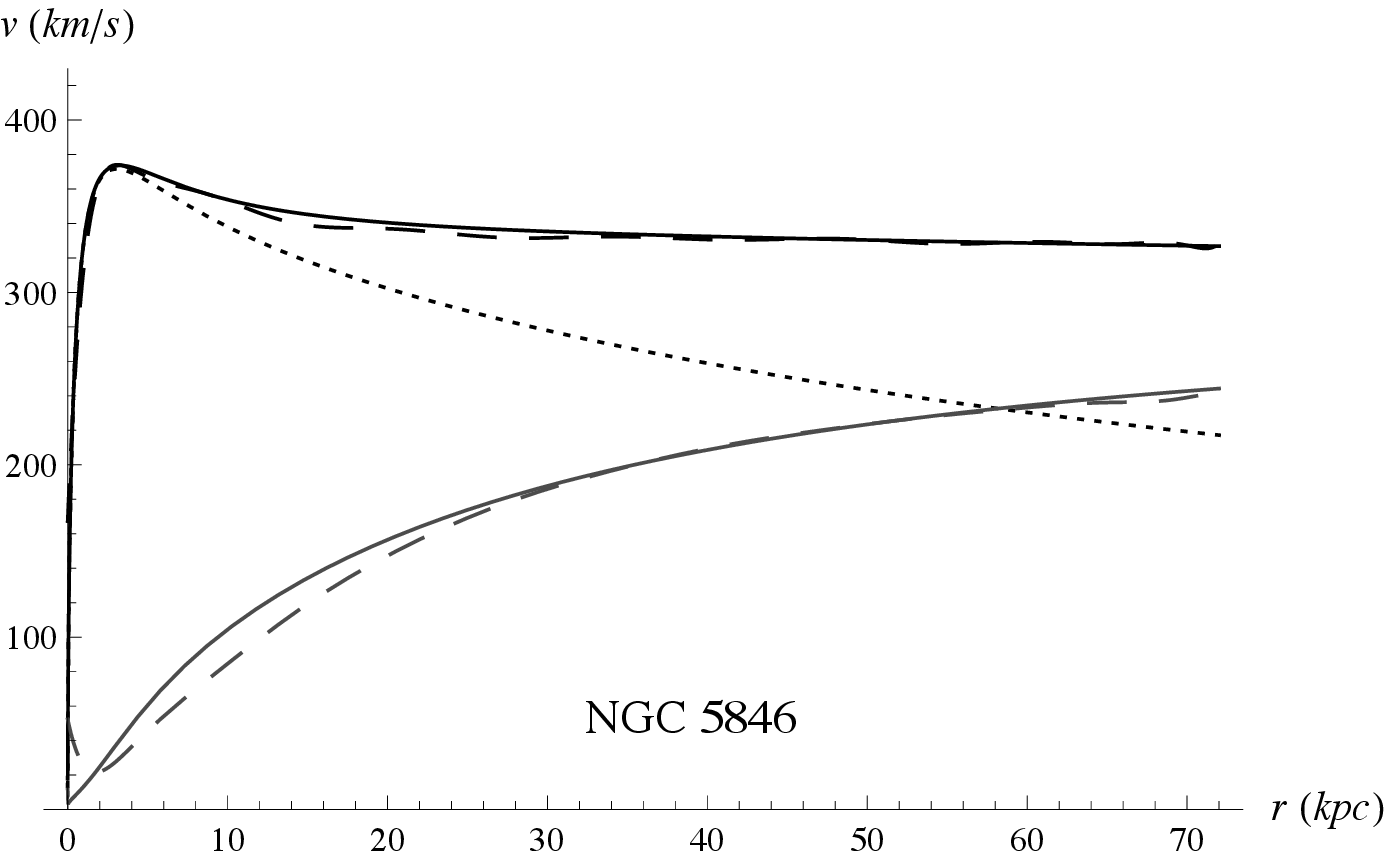} \\
	
	\includegraphics[width=7.5cm]{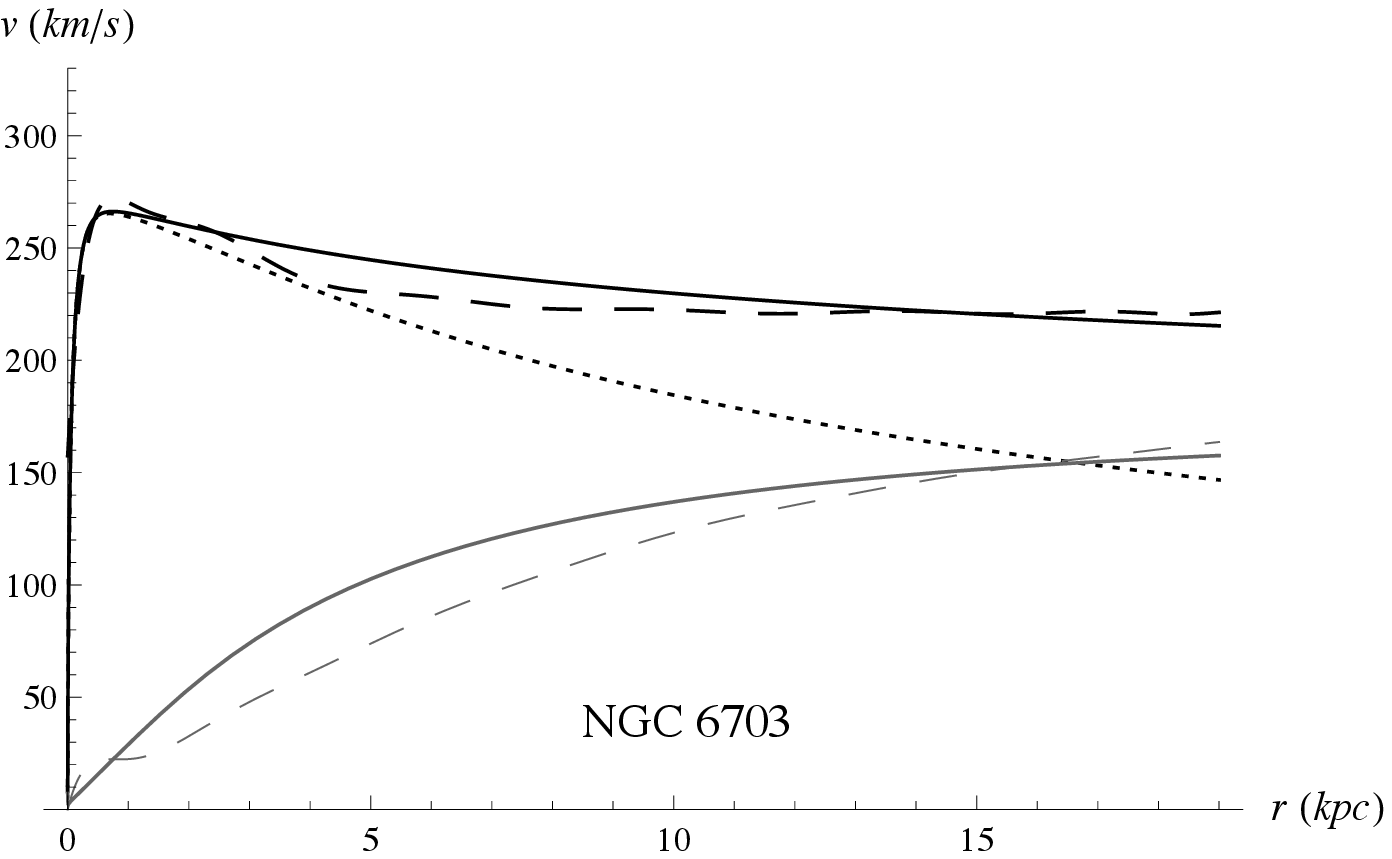} &
	\includegraphics[width=7.5cm]{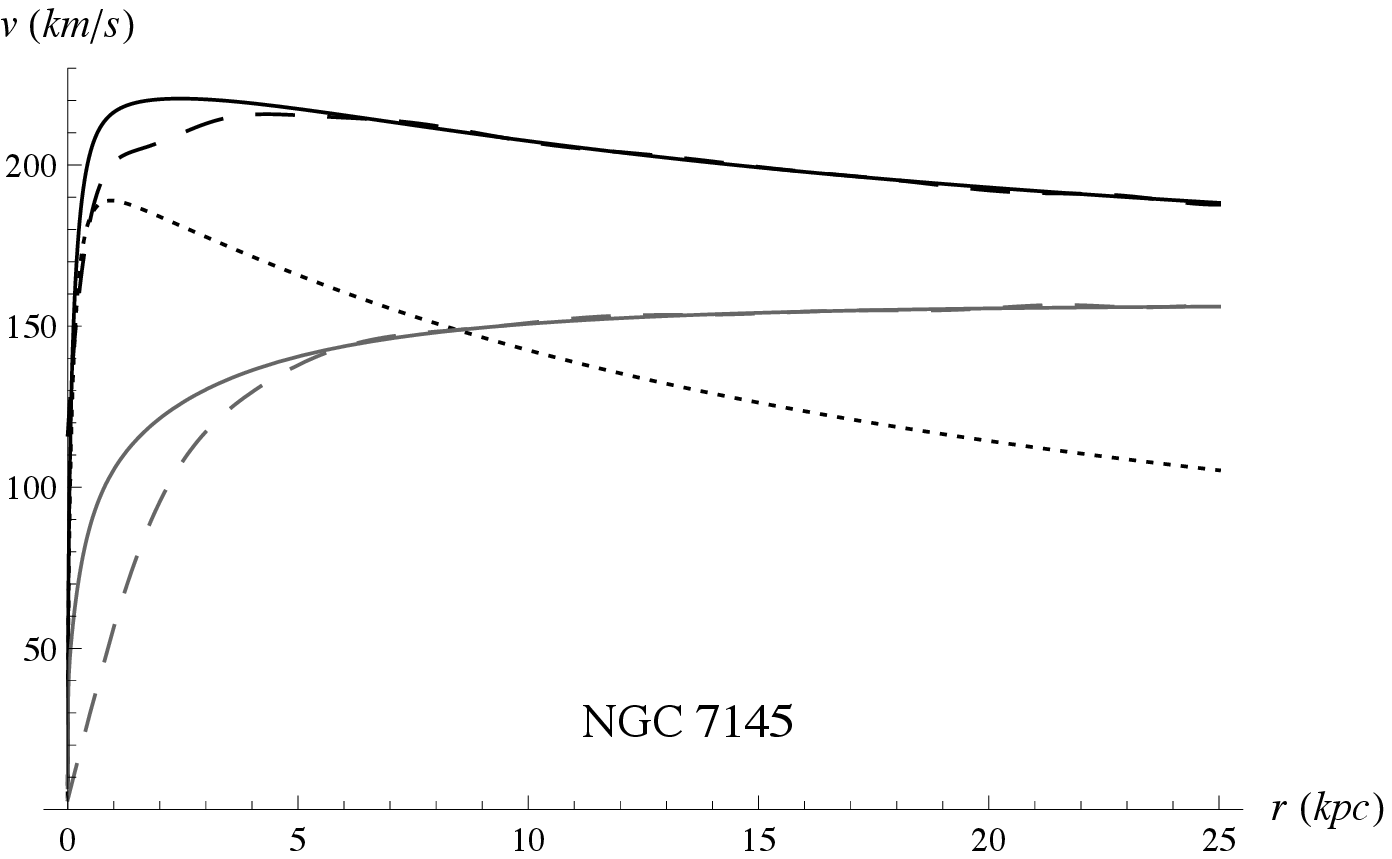} \\

	\includegraphics[width=7.5cm]{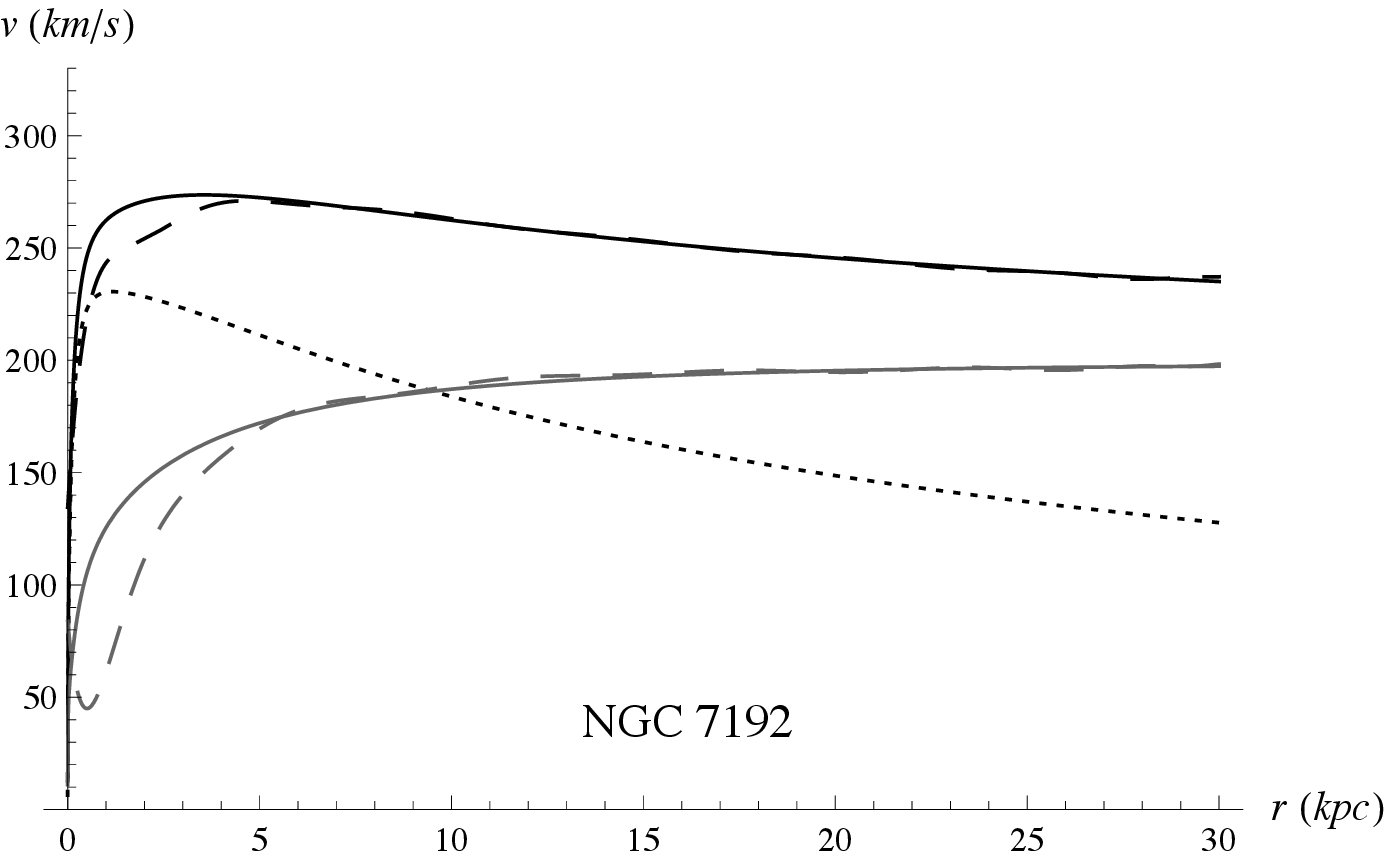} &
	\includegraphics[width=7.5cm]{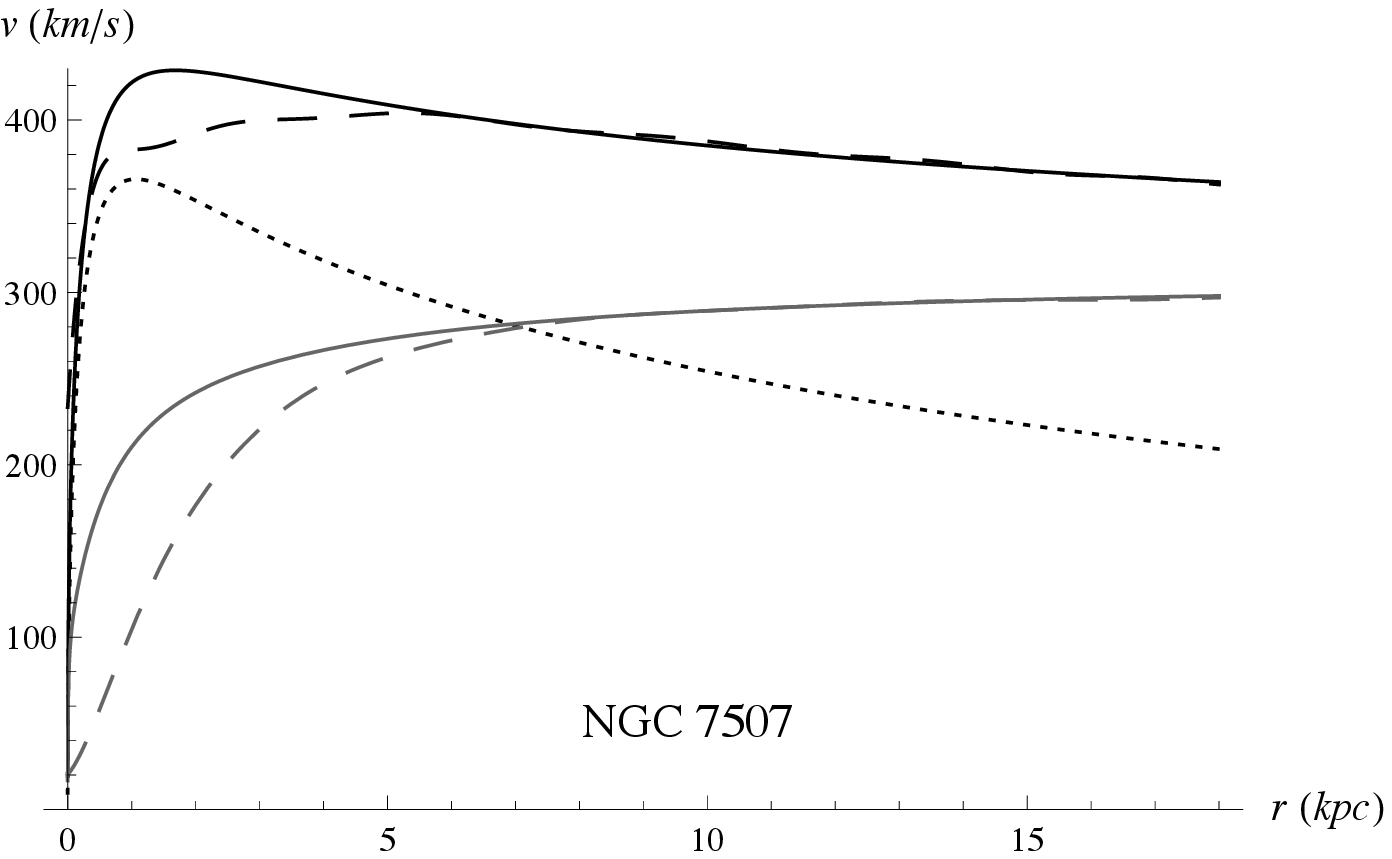} \\
	
	\includegraphics[width=7.5cm]{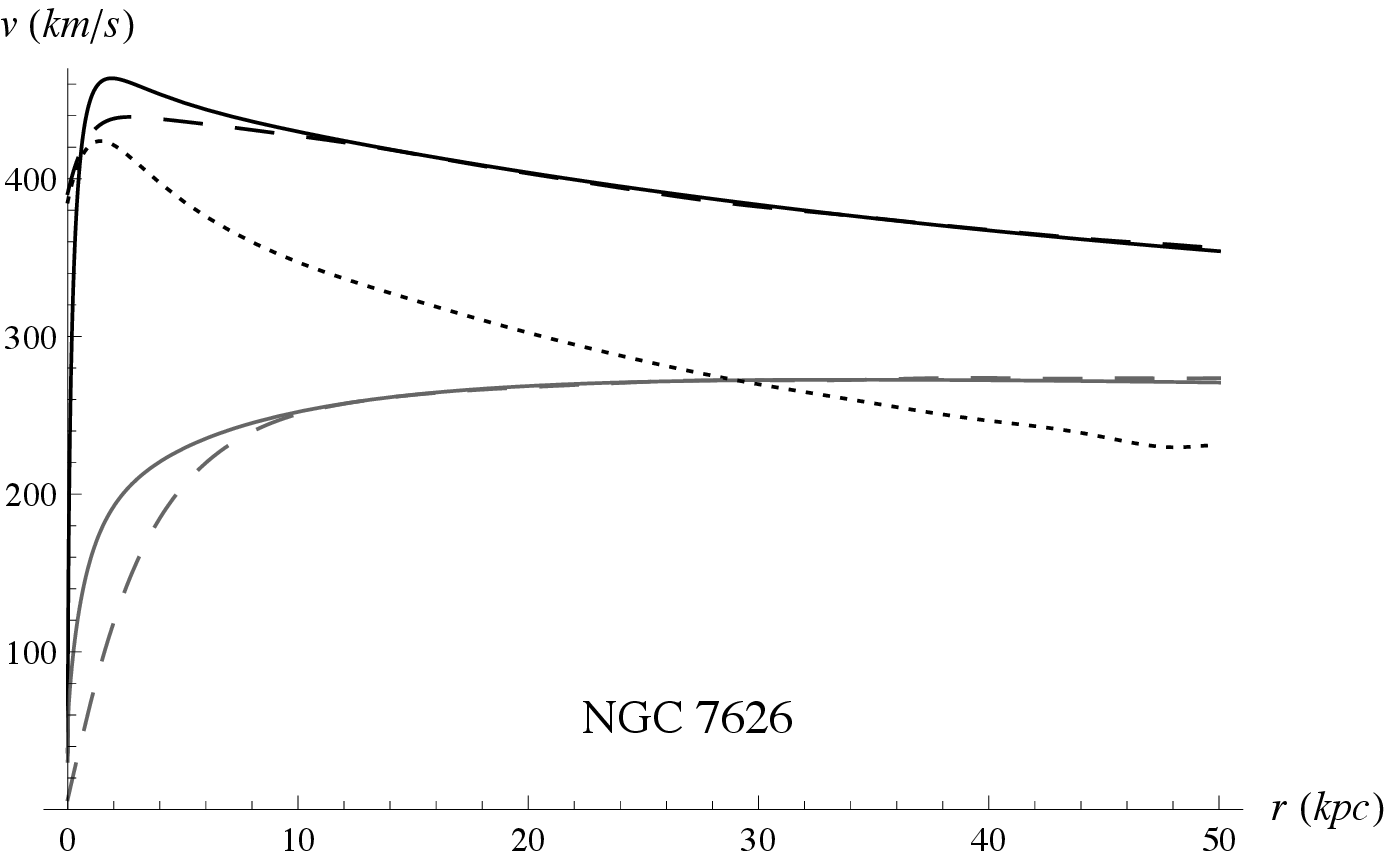}

\end{array}$
\end{center}
\caption{{\bf Left}: Observed rotation curve (dashed full), decomposed into visible (dotted) and dark matter (dashed grey) contributions  \cite{kronawitter}, superimposed with the mimicked dark matter profile (full grey) arising from the composite non-minimal coupling and resulting full rotation curve (full).
{\bf Right}: Log-Log profile of the visible matter density (dotted), mimicked ``dark matter'' contribution (dashed) and the sum of both components (full).}

\label{fig1}
\end{figure}

The results derived from the numerical solution of \eq{tracetwo} are depicted in Table \ref{tableboth}, and shown in Fig.\ref{fig1}: as can be seen, the composite non-minimal coupling \eq{f2two} provides close fits for all rotation curves in the outer region --- of much higher quality than those obtainable if one assumed individual couplings $f_2(R) = R/R_1$ or $f_2(R) = \sqrt[3]{R/ R_3}$ (see additional graphs in Ref. \cite{paramos}). For some galaxies, the overall quality of fit is disturbed by a discrepancy in the inner galactic region. This could result from an uncertainty in the original derivation of the rotation curves, or perhaps due to the deviation from purely spherical symmetry.

From a fundamental standpoint, this deviation could also indicate that the non-minimal coupling should include an additional term that acquires relevance in a higher density environment: recalling an earlier work in the high density context of the Sun, one could admit a linear addition $\de f_2(R) \propto R$, for instance. Of course, the considered model could be degenerate to the approximation of the pure curvature term given by the GR prescription $f_1(R) = 2\ka R$, and a more complex picture involving both non-trivial $f_1(R)$ and $f_2(R)$ couplings is in order.

Although details are here suppressed, for brevity, after obtaining the solution to \eq{tracetwo} it is straightforward to compute the gradient term on its {\it r.h.s.}: as discussed before, this is indeed dominant, so that the mimicked dark matter density profile is almost completely overlapped with it.


\begin{table}
\begin{center}
\begin{tabular}[h]{c|cc|cc|cc}

Galaxy & \multicolumn{4}{c|}{Composite} & \multicolumn{2}{c}{Single} \\
(NGC)& $r_1$ & $r_3$  &  $r_{\infty ~1 }$  & $r_{\infty ~3 }$   &  $r_1$ & $r_3$  \\ 
\hline 
2434 &  $\infty$ & 0.9&   0 & 33.1   &   4.1 & 0.9 \\
5846 & 37 & $\infty $ &   138 & 0    &  37 & 34.9 \\
6703 &  22 & $\infty$   &  61.2 & 0    &  22 & 26.2 \\
7145 & 22.3 & 47.3  &  60.9 & 14.2   &   19.9 & 23.8  \\
7192 &  14.8 & 24   &  86.0 & 18.3    &   14.5 & 7.3 \\
7507 &  4.9 & 2.9  &  178 & 31.1   &   4.3 & 1.1 \\
7626 &  28 & 9.6  &  124 & 42.5   &  16.0 & 7.1 

\end{tabular}
      
\caption{Best fit values for the characteristic lengthscales $r_1$ and $r_3$ for the composite and separate fits of the galaxy rotation curves with the $n=-1$ isothermal sphere and $n=-1/3$ NFW mimicked ``dark matter'' scenarios, together with background matching distances $r_{\infty~1}$ ($n=-1$) and $r_{\infty ~3}$ ($n=-1/3$) for the composite non-minimal coupling. The limit $ r_i = \infty$ for the composite scenario indicates that the corresponding scale $R_i = 0$ . The units used are $r_1$ ($Gpc$), $r_3$ ($10^5~Gpc$), $r_{\infty~1}$ and $r_{\infty~3}$ ($kpc$). \label{tableboth}}

\end{center}
\end{table}



\subsection{Characteristic density, background matching and mimicked ``dark matter'' inter-dominance}

The selected galaxies have their rotation curves fitted by characteristic lengthscales of the order $r_1 \sim 10~Gpc$ and $r_3 \sim (10^5-10^6)~Gpc$. Although the analytic results derived in the preceding sections do not hold strictly in the assumed composite coupling scenario of \eq{f2two}, \eq{matchingr} should nevertheless provide a quantitative figure for the background matching distances,

\beqa r_\infty &=& \left[{2\over 3} \times 3^n (1-2n) 
\right]^{1/4} \left( {c \over r_0 H } \right)^{(1-n)/2} \left( r_s r_0^2  a\right)^{1/4}  \nonumber \approx \\ 
&\simeq &\cases{
\left( {3.8~Gpc \over r_3} \right)^{2/3} \left( r_s r_3^2  a\right)^{1/4}   & , $n = -1/3$
\cr \left( {3.8~Gpc \over r_1} \right) \left( r_s r_1^2  a\right)^{1/4}  &,  $n=-1$
}, \eeqa

\noindent after inserting the Hubble constant $H_0 = 70.8 ~km ~s^{-1}~Mpc^{-1}$. Notice that this 
expression was obtained with the assumption of a single Hernquist profile for the visible matter density, and not for the two component 
model used in the current fitting session; using the dominant contribution stemming from the extended gas distribution, one obtains the values displayed in the last columns of Table \ref{tableboth}. 

Since $r_\infty$ may be regarded as signaling the radius of the mimicked dark matter haloes, Table \ref{tableboth} shows that the $n=-1/3$ NFW haloes are always smaller than those with the $n=-1$ isothermal sphere profile --- with the exception of NGC 2434 that, however, stands out from the set since it is best fitted by a single NFW ``dark matter'' component. Also, $r_{\infty~1}$ is always larger than the range of the rotation curves, so that the isothermal sphere ``dark matter'' extends further than the observationally inferred; on the contrary, the value of $r_{\infty~3}$ goes from about $ 0.57 L$ (NGC 7145) to $1.73 L$ (NGC 7507), $L$ being the full distance considered for the rotation curves.

This indicates that the isothermal sphere halo is more significant that its NFW equivalent: even the exceptional case of NGC 2434, where no isothermal sphere component appears in the best fit scenario, yields a matching distance $r_{\infty~3} = 33.1~kpc$ --- only about twice the endpoint of the corresponding rotation curve, $L \simeq 17~kpc$.

This is confirmed by a simple algebraic computation: since the NFW profile falls as $r^{-3}$, while the isothermal sphere one displays a shallower $r^{-2}$ long-range dependence, and the respective characteristic densities obey $\rho_3 \ll \rho_1$, it is easy to conclude that the isothermal sphere eventually becomes the dominant dark matter contribution. Indeed, equating both contributions yields 

\beq {\rho_{dm~3} \over \rho_{dm~1}} = \left({r_1^2 \over a r_3 }\right)^{1/2} \left({r_s\over a}  \right)^{1/4} \left({a \over r}\right) \sim {1 
\over 10} {a \over r}  ,\eeq

\noindent so that this dominance occurs at distances larger than $r = a/10 \sim 1~kpc$. Thus, the considered galactic rotation curves are found to be ``mostly flat''.

\subsection{Cosmological relevance}

As discussed already, the results obtained above are naturally tuned towards the corresponding astrophysical setting, if one assumes that the chosen form for the (composite) non-minimal coupling represents the leading terms of a series expansion of a more evolved expression for $f_2(R)$, suitable in the mentioned range for the scalar curvature $R$. Hence, there is nothing mandatory concerning a possible cosmological relevance of the composite coupling \eq{f2two}, with the obtained values for $R_1 $ and $R_3$: it might be the case that a suitable mechanism applicable in cosmology corresponds to another term of the hypothetical series expansion of $f_2(R)$.

Nevertheless, it is natural to assess if the considered cases $n= -1$ and $n=-1/3$ give rise to implications for cosmology, since this would present a natural candidate for a unification model of dark matter and dark energy. From the discussion of paragraph \ref{matching}, one may write the non-
minimal coupling \eq{f2two}, in a cosmological context, as

\beq f_2 = \sqrt[3]{R_3 \over R} + {R_1 \over R} \simeq \left( { r_H \over r_3} \right)^{2/3} + \left( { r_H \over r_1} \right)^2 \sim 10^{-4} + 
10^{-1} , \eeq

\noindent having used $R = 3/r_H^2$, with $r_H = c /H = 4.2~Gpc$ the Hubble radius, and inserting the best fit orders of magnitude $r_1 \sim 10~Gpc$ and $r_3 \sim (10^5-10^6)~Gpc$.

The small value of the $n=-1/3$ non-minimal coupling indicates that this should not be significant at a cosmological level; the higher value of the $n=-1$ power-law coupling (when compared to unity) hints that a simple $R_1/R$ coupling could prove interesting for cosmology.

\subsection{Universality}

The obtained values for $r_1$ and $r_3$, shown in Table \ref{tableboth},  display an undesired variation between galaxies, since these parameters should be universal. Indeed, $r_1$ averages $\bar{r}_1=21.5 ~Gpc$ with a standard deviation $\si_1=10.0~Gpc$, while $r_3$ 
presents $\bar{r}_3 = 1.69 \times 10^6~Gpc$ and $\si_3 = 1.72 \times 10^6 ~Gpc$. This deviation from universality is not unexpected, and several causes could be hinted as culprits: although small, the existing assymetry in the selected type E0 galaxies due to the $1 - b/a \lesssim 0.1$ semi-axis deviation from pure sphericity could translate into the difference in obtained values for $r_1$ and $r_3$. This shift from the assumed symmetry could be enhanced by localized features and inhomogeneities, which could also act as ``seeds'' for the dynamical generation of the mimicked ``dark matter'' component. Furthermore, the reported dominance of the gradient term in the {\it r.h.s.} of \eq{tracetwo} could boost this effect, if these perturbations vary significantly on a short lengthscale.

By the same token, the relevance of the gradient term could lead to the deviation from universality if the chosen profiles (Hernquist for visible matter, NFW and isothermal sphere for mimicked dark matter) prove not to be so suitable as hoped: this mismatch between the real density profiles and the fitted ones does not need to be large, if  a large deviation of its first and second derivatives lies in very localized regions: this could act as a sort of ``fitting inhomogeneities'' with the same effect as of those discussed above.

Aside from the symmetry and fitting functions selected, more fundamental issues could be behind  the variation of the obtained values for $r_1$ and $r_3$: in fact, the non-minimal coupling \eq{f2two} could be too simplistic, and a more complex model could provide a more universal fit of the galaxy rotation curves.

Also (as discussed with respect to the relatively less successful fits of the inner galactic regions), the simplifying assumption of a linear curvature term $f_1(R) = 2\ka R$ might also contribute to this deviation from universality, since one knows that power-law terms $f_1(R) \propto R^n$ can also induce a dynamical ``dark matter'' \cite{CapoLSB}. 
 
Finally, a more interesting idea could lie behind the obtained discrepancy, namely that the visible matter Lagrangian density itself may be different from the assumed perfect fluid form, ${\cal L}_m = -\rho$ \cite{fluid,fluidfaraoni}: this functional dependence is inadequate in the context of the model here addressed, and a more adequate form ${\cal L}_m = {\cal L}_m(\rho)$ should be considered --- perhaps even including other variables related to the thermodynamical description of the matter distribution (see Ref. \cite{harko} for an in-depth discussion based upon a similar consideration).

\section{Conclusions}

In this work the possibility of obtaining a solution to the dark matter puzzle, embodied by the flattening of 
galaxy rotation curves, was approached by resorting to the main phenomenological implications of models possessing a non-minimal 
coupling of matter to curvature, following results reported in Ref. \cite{paramos}.

As a first attempt, one first examined the non-conservation of the energy-momentum tensor of matter and the implied deviation from geodesic motion, and ascertained what coupling $f_2(R)$ should be so that the derived extra force would lead to the reported flattening of the rotation curves. This requires a logarithmic coupling of the form $\la f_2(R) = -v^2/m \log (R/
R_*)$ (where $m$ is the outer slope of the visible matter density $\rho$), which can be approximated by a simpler power-law $\la f_2(R) \approx (R_*/R)^\al$, with the asymptotic velocity given by $v_\infty^2 = \al m$. However, this solution yields an almost universal $v_\infty$, prompting for another possible mechanism for the dynamical mimicking of ``dark matter''.

This need led to the second, more evolved approach, which is rooted empirically in the phenomenological Tully-Fisher relation: one may instead assume that geodesical motion is preserved, $\nabla^\mu T_\mn = 0$ (a condition proved to be self-consistent), but that the metric itself is perturbed: the mimicked dark matter density is then given by the difference $R/2\ka -  \rho$. By resorting to a power-law coupling with matter $f_2(R) = (R/R_0)^n$ (with a negative exponent $n$ yielding the desired effect at low curvatures and long range).

The proposed power-law directly leads to a ``dark matter'' density that depends on a power of the visible matter density, thus accounting for the Tully-Fisher law in a natural way. The obtained ``dark matter'' component has a negative pressure, as commonly found in cosmology in dark energy models: this hints the possibility that the non-minimal coupling model might unify dark matter and dark energy.

Given their relevance in the literature, two different scenarios were considered: the NFW and isothermal sphere dark matter profiles ($n=-1/3$ 
and $n=-1$, respectively). Although separate fits of the selected galaxy rotation curves to each coupling do not yield satisfactory results, a composite coupling of both power-laws produced a much improved adjustment.

The characteristic lengthscales $r_1$ and $r_3$ were taken as fitting parameters for each individual galaxy, yielding the order of magnitudes $r_1 \sim 10~Gpc$ and $r_3 \sim 10^5 ~Gpc$. This enabled the computation of the cosmological background matching distances for each galaxy (which depend on their characteristic lengthscale $a$) and the obtained astrophysical range showed that the $n=-1$ isothermal sphere ``dark matter'' halo dominates the $n=-1/3$ NFW component. Furthermore, the $n=-1$ scenario was shown to have possible relevance in a  cosmological context, as $r_1 \sim r_H$, the latter being the Hubble radius --- again hinting at a possible unification of the dark components of the Universe.

The lack of the desirable universality in the model parameters $R_1$ and $R_3$ was duely noticed, although these quantities are characterized by the same order of magnitude for the best fits obtained; several possible causes for this variation were put forward, from deviation from spherical symmetry to a more complex form for the non-minimal coupling between curvature and matter $f_2(R)$ or the curvature term $f_1(R)$, or the need for a more evolved Lagrangian description of the latter.

As a final remark, one concludes that the rich phenomenology that springs from the model with a non-minimal coupling between matter and curvature enables a direct and elegant alternative to standard dark matter scenarios and many modifications to GR, which usually resort to extensive use of additional fields and other {\it ad-hoc} features. As an example of the latter, one points out the MOdified Newtonian Dynamics (MOND) hypothesis, which is by itself purely phenomenological, and whose underlying Tensor-Vector-Scalar theory is based upon an extensive paraphernalia of vector and scalar fields \cite{MOND1,MOND2,MOND3,MOND4} (see Ref. \cite{MONDcritic} for a critical assessment). As a final remark, we point out that the considered model may also be translated into a multi-scalar theory \cite{scalar}, with two scalar fields given by

\beq \varphi^1 = {\sqrt{3}\over2} \log \left[ 1 + n \left({R \over R_0}\right)^n  \right] \qquad \varphi^2 = R, \eeq 

\noindent with dynamics driven by a potential

\beq U(\varphi^1,\varphi^2) =  {1 \over 4} \exp \left( -{2 \sqrt{3}\over 3} \varphi^1 \right) \left[\varphi^2 - 
{f_1(\varphi^2 )\over 2\ka }  \exp \left( -{2 \sqrt{3}\over 3} \varphi^1 \right)  \right]. \eeq

\ack The authors thank E. Vagena for fruitful discussions and the hospitality shown during the {\it First Mediterranean Conference on Classical and Quantum Gravity}, Crete, September 2009.
\par ~

\bibliography{crete}{}
\bibliographystyle{iopart-num}

\end{document}